\documentclass[aps]{revtex4}
\usepackage{epsfig}
\usepackage{amsbsy}
\usepackage{amsmath}
\usepackage{amsfonts}
\usepackage{graphicx}

\newcommand{\nn}{\nonumber}
\newcommand{\ip}[1]{\left\langle{#1}\right\rangle}

\newcommand{\ket}[1]{|{#1}\rangle}
\newcommand{\braket}[2]{\langle{#1}|{#2}\rangle}

\begin{document}
\date{\today}
\title{The effects of time delays in adaptive phase measurements}
\author{D. W. BERRY$^{1}$ AND H. M. WISEMAN$^{2,1}$ \\
$^{1}$Department of Physics, The University of Queensland, St.\ Lucia
4072, Brisbane, Australia \\
$^{2}$School of Science, Griffith University, Nathan 4111, Brisbane,
Australia}

\begin{abstract}
It is not possible to make measurements of the phase of an optical
mode using linear optics without introducing an extra phase
uncertainty. This extra phase variance is quite large for heterodyne
measurements, however it is possible to reduce it to the theoretical
limit of $\log \bar n/(4\bar n^2)$ using adaptive measurements. These
measurements are quite sensitive to experimental inaccuracies,
especially time delays and inefficient detectors. Here it is shown that
the minimum introduced phase variance when there is a time delay of
$\tau$ is $\tau/(8\bar n)$. This result is verified numerically, showing
that the phase variance introduced approaches this limit for most of
the adaptive schemes using the best final phase estimate.
The main exception is the adaptive mark II scheme with simplified
feedback, which is extremely sensitive to time delays. The extra phase variance
due to time delays is considered for the mark I case with
simplified feedback, verifying the $\tau /2$ result obtained by Wiseman and
Killip both by a more rigorous analytic technique and numerically.
\end{abstract}
\maketitle

\renewcommand{\thesection}{\arabic{section}}
\section{Introduction}
It is well known that any measurement of the phase of an
electromagnetic field using linear optics will have an uncertainty
that is greater than the intrinsic quantum phase uncertainty of the
state \cite{fullquan}. The standard phase measurement method is the
heterodyne scheme, where the signal is combined with a strong local
oscillator field with a slightly different frequency. This means that
all quadratures of the field are sampled approximately equally. This
phase measurement method introduces a phase variance of approximately
$1/(4\bar n)$, where $\bar n$ is the mean photon number of the field.

This is not so significant for measurements on coherent states, where
it is the same size as the intrinsic phase uncertainty of the state.
For states with reduced phase uncertainty, however, the uncertainty
in the measurement will be far greater than the intrinsic phase uncertainty.
For example, minimum phase uncertainty states have a phase variance
that scales as $1/(\bar n^2)$ \cite{SumPeg90}.

It is possible to dramatically improve on heterodyne measurements by
using a local oscillator phase of $\Phi=\varphi+\pi/2$, where $\varphi$
is the signal phase (homodyne detection). This has the drawback that
the phase must be known in advance. Adaptive phase measurements
\cite{fullquan,Wis95c,semiclass,BerWisZha99,unpub} attempt
to approximate a homodyne phase measurement by adjusting the local oscillator
phase based on data obtained during the measurement.

There are several different variations of adaptive phase measurements.
The adaptive mark I scheme gives improved results for small photon numbers,
but worse results for large photon numbers. The adaptive
mark II scheme gives an introduced phase variance of $1/(8\bar n^{1.5})$
\cite{semiclass,BerWisZha99}, a significant improvement over heterodyne
measurements. By a more sophisticated feedback algorithm it is even possible
to obtain the theoretical limit of $\log {\bar n}/(4\bar n^2)$ \cite{unpub}.

These adaptive measurement schemes are sensitive to experimental
imperfections, most notably imperfect detectors and time delays. The
effect of imperfect detectors is fairly straightforward, introducing a
phase variance of approximately $(1-\eta)/(4\eta\bar n)$ \cite{semiclass}.
The effect of time delays is more difficult to estimate. Some highly
simplified calculations indicate that the excess phase variance due to
time delays is $\tau/2$ for mark I measurements (where $\tau$ is the time
delay), and $\tau/(2\bar n)$ for mark II measurements \cite{semiclass}.

Here we repeat these derivations more accurately, and show that while
the result for mark I measurements is reasonably accurate, the
perturbation approach is inadequate to obtain a consistent result for
mark II measurements. We consider an alternative derivation that gives
the minimum phase variance when there is a time delay. In
section~\ref{result} we evaluate the phase variance with time delays
numerically and show that for most of the measurement schemes the
phase variance approaches this limit for large time delays.

\section{Background theory}
Before we proceed to determining the effect of time delays, we will
briefly outline the background theory for adaptive measurements. For
more details see \cite{fullquan,semiclass}. First the
photocurrent for dyne detection \cite{fullquan} is given by
\begin{equation}
I(v){\rm d}v=2{\rm Re}(\alpha_v e^{-i\Phi(v)}){\rm d}v+{\rm d}W(v).
\end{equation}
Here $\alpha_v$ is the scaled coherent amplitude of the signal, $v$ is
the scaled time and $\Phi(v)$ is the local oscillator phase. The
systematic variation with time of the coherent amplitude due to the
mode shape is scaled out, and time is scaled to the unit interval. We
also define the variables
\begin{align}
A_v &= \int_0^v I(u)e^{i\Phi(u)}{\rm d}u ,\\
B_v &= -\int_0^v e^{2i\Phi(u)}{\rm d}u,\\
C_v &= A_v v+B_v A_v^*.
\end{align}
We omit the subscripts to indicate final values. The best estimate of
the phase at time $v$ is given by $\arg C_v$, and if $|B_v|$ is small
then $\arg A_v$ is also a good phase estimate.

For mark I measurements $\arg A_v$ is used as the intermediate phase
estimate (so the local oscillator phase is 
$\Phi = \arg A_v + \pi/2$) and also as the phase estimate at the
end of the measurement. For mark II measurements we use the same
intermediate phase estimate but we use the best estimate $\arg C$ at
the end of the measurement. In \cite{semiclass} it is shown that using
$\arg A_v$ as the intermediate phase estimate is equivalent to varying
the local oscillator phase as
\begin{equation}
\label{simp}
{\rm d}\Phi(v)=\frac{I(v){\rm d}v}{\sqrt v}.
\end{equation}
Using the feedback in this form allows us to use a simplified feedback
circuit experimentally. This feedback is no longer equivalent to using
a phase estimate of $\arg A_v$ when there are time delays in the system,
and we therefore consider the cases with $\arg A_v$ feedback and
simplified feedback separately.

The next feedback scheme uses a phase estimate intermediate between
$\arg A_v$ and $\arg C_v$, specifically
\begin{equation}
\hat \varphi = \arg(A_v^{\epsilon}C_v^{1-\epsilon}).
\end{equation}
The simplest case is where $\epsilon$ does not depend on time. For each
mean photon number there is an optimum value to use, and provided these
optimum values are used this method gives better results than mark II
measurements \cite{unpub}.

We can get very close to the theoretical limit of $\log \bar n/(4\bar n^2)$
if we use a time dependent $\epsilon$ given by \cite{unpub}
\begin{equation}
\label{vareps}
\epsilon(v) = \frac{v^2-|B_v|^2}{|C_v|}\sqrt{\frac v{1-v}}.
\end{equation}
Briefly explaining the reason for the theoretical limit, the probability
distribution for $A$ and $B$ is proportional to
$|\braket{\beta,\zeta}{\psi}|^2$, where $\ket{\psi}$ is the signal state
and $\ket{\beta,\zeta}$ is a squeezed state
\begin{equation}
\label{alze}
\ket{\beta,\zeta}=\exp(\beta a^\dagger-{{\beta}}^*a)
\exp[\tfrac{1}{2} \zeta^* a^2-\tfrac{1}{2} \zeta (a^\dagger)^2]\ket{0},
\end{equation}
where
\begin{align}
\beta&=\frac{C}{1-B^2}, \\
\zeta&=-\frac{B {\rm atanh} |B|}{|B|}.
\end{align}
This means that the phase variance introduced is approximately equal to
the phase variance of the squeezed state $\ket{\beta,\zeta}$. The photon
number of this squeezed state will be approximately the same as that of
the input state, so the theoretical minimum introduced phase variance is
that of an optimized squeezed state with photon number $\bar n$. As shown
in \cite{collett}, this scales as $\log{\bar n}/\bar n^2$.

The phase measurement scheme with time dependent $\epsilon$ (\ref{vareps})
is not quite at the theoretical limit because it produces squeezed states
$\ket{\beta,\zeta}$ that are slightly more highly squeezed than optimum.
In \cite{unpub} we show how this can be corrected, but we will not
consider the case with these corrections here because the time delays
cause the squeezed state to be less squeezed than optimum anyway, so these
corrections are not needed.

\section{Perturbation approach}
\subsection{Mark I}
\label{dtontwo}
Now we estimate the effect of time delays on simplified mark I
measurements in a similar way to that done in \cite{semiclass}, but using
fewer of the simplifications used there. Without a time delay the
stochastic differential equation (SDE) for the phase estimate is
\begin{align}
{\rm d}\hat \varphi_v &= \frac{I(v) {\rm d}v}{\sqrt v} \nonumber \\
&= v^{-1/2} [-2\alpha \sin \hat \varphi_v {\rm d}v + {\rm d}W(v)].
\end{align}
In this expression we have taken the input phase to be zero. For some
time $v_1$ the phase will come to lie near 0, so we linearise around
$\hat \varphi_v = 0$.  The result, which will be valid for
$v_1 \le v \le 1$ is
\begin{equation}
{\rm d}\hat \varphi_v = v^{-1/2} [-2\alpha \hat \varphi_{v} {\rm d}v +
{\rm d}W(v)].
\label{nodelay}
\end{equation}
Including the time delay the SDE is
\begin{equation}
{\rm d}\hat \varphi_v = v^{-1/2} [-2\alpha \hat \varphi_{v-\tau} {\rm d}v +
{\rm d}W(v)].
\end{equation}
Now we treat the delay perturbatively.  We write the solution to the
perturbed equation as
\begin{equation}
\hat\varphi_v=\hat\varphi_v^{(0)}+\alpha\tau\hat\varphi_v^{(1)}
+O(\alpha^2\tau^2).
\end{equation}
The zeroth-order term obeys the SDE for no delay (\ref{nodelay}), so the
first-order correction obeys
\begin{equation}
\alpha\tau {\rm d}\hat \varphi_v^{(1)} = 2\alpha v^{-1/2}(
\hat\varphi_v^{(0)}-\hat\varphi_{v-\tau}^{(0)}){\rm d}v-2\alpha^2\tau 
v^{-1/2}\hat\varphi_{v-\tau}^{(1)}{\rm d}v.
\end{equation}
Therefore to first order in $\tau$ we have
\begin{align}
{\rm d}\hat \varphi_v^{(1)} &= 2 v^{-1/2} {\rm d}\hat\varphi_v^{(0)}-2\alpha
v^{-1/2}\hat\varphi_v^{(1)}{\rm d}v \nn \\
&=2 v^{-1/2} \{v^{-1/2} [-2\alpha \hat\varphi_v^{(0)} {\rm d}v + {\rm d}W(v)]\}
-2\alpha v^{-1/2}\hat\varphi_v^{(1)}{\rm d}v \nn \\
&=-2\alpha v^{-1/2}\hat\varphi_v^{(1)}{\rm d}v-\frac {4\alpha}v \hat\varphi_v^
{(0)} {\rm d}v +\frac 2v {\rm d}W(v).
\label{firstorder}
\end{align}
It is straightforward to show that the solution to the zeroth-order
equation is
\begin{equation}
\hat \varphi_v^{(0)}=e^{4\alpha(\sqrt {v_1}-\sqrt v)}\hat\varphi_{v_1}^{
(0)}+\int_{v_1}^v \frac {e^{4\alpha(\sqrt u-\sqrt v)}} {\sqrt u} {\rm d}W(u).
\end{equation}
Using this in equation (\ref{firstorder}) and multiplying
on both sides by $e^{4\alpha\sqrt v}$ gives
\begin{equation}
{\rm d}(e^{4\alpha\sqrt v}\hat \varphi_v^{(1)}) = -\frac{4\alpha}v \left[
e^{4\alpha\sqrt {v_1}}\hat\varphi_{v_1}^{(0)}+\int_{v_1}^v \frac
{e^{4\alpha\sqrt u}} {\sqrt u} {\rm d}W(u) \right] {\rm d}v
+\frac 2v e^{4\alpha\sqrt v} {\rm d}W(v).
\end{equation}
Integrating then gives the solution
\begin{align}
\hat \varphi_v^{(1)} &= e^{4\alpha(\sqrt {v_1}-\sqrt v)}
\hat \varphi_{v_1}^{(1)} -\int_{v_1}^v\frac{4\alpha}s \left[ e^{4
\alpha(\sqrt{v_1}-\sqrt v)}\hat\varphi_{v_1}^{(0)} \vphantom{\frac {e^{4\alpha
(\sqrt u-\sqrt v)}}{\sqrt u}}\right. \nn \\ & \left. +\int_{v_1}^s \frac
{e^{4\alpha(\sqrt u-\sqrt v)}}{\sqrt u}{\rm d}W(u) \right] {\rm d}s +
\int_{v_1}^v\frac 2s e^{4\alpha(\sqrt s-\sqrt v)}{\rm d}W(s).
\end{align}
In this approximation the mark I phase estimate is given by
$\hat\varphi_1=\hat\varphi_1^{(0)}+\alpha\tau\hat\varphi_1^{(1)}$. To
first order in $\tau$ and $\alpha^{-1}$ this has a variance of
\begin{equation}
\ip{\phi_{\rm I}^2}=\frac 1{4\alpha}+2\alpha\tau\ip{\hat\varphi_1^{(0)}
\hat\varphi_1^{(1)}}.
\end{equation}
Here we have used the known variance of $1/(4\alpha)$ of the zeroth-order
term. Evaluating the second term on the right-hand side we find
\begin{align}
\ip{\hat\varphi_1^{(0)}\hat\varphi_1^{(1)}} &= e^{-8\alpha(1-\sqrt{v_1})}\ip{
\hat\varphi_{v_1}^{(0)}\hat\varphi_{v_1}^{(1)}} + 4\alpha\log(v_1)
e^{-8\alpha(1-\sqrt{v_1})}\ip{{\hat{\varphi}_{v_1}^{(0)2}}} \nn \\
&-\int_{v_1}^1 \frac {4\alpha}s \int_{v_1}^s \frac {e
^{8\alpha(\sqrt u-1)}}u {\rm d}u {\rm d}s +\int_{v_1}^1 \frac {2 e^{8\alpha
(\sqrt u-1)}}{u^{1.5}}{\rm d}u.
\end{align}
The first two terms decrease exponentially with $\alpha$ and may
therefore be omitted. Exchanging the order of the integrals in the
third term and integrating gives
\begin{equation}
\ip{\hat\varphi_1^{(0)}\hat\varphi_1^{(1)}} = 4\alpha\int_{v_1}^1 \frac {\log u}
u e^{8\alpha(\sqrt u-1)} {\rm d}u + \int_{v_1}^1 \frac {2 e^{8\alpha
(\sqrt u-1)}}{u^{1.5}}{\rm d}u.
\end{equation}
Now we change variables to $s=1-\sqrt u$, so ${\rm d}u=-2(1-s){\rm d}s$. Then we
obtain
\begin{equation}
\ip{\hat\varphi_1^{(0)}\hat\varphi_1^{(1)}} = 16\alpha\int_0^{1-\sqrt{v_1}}
\frac {\log (1-s)} {(1-s)} e^{-8\alpha s} {\rm d}s +
4\int_0^{1-\sqrt{v_1}} \frac {e^{-8\alpha s}}{(1-s)^2}{\rm d}s.
\end{equation}
Expanding in a Maclaurin series in $s$ gives
\begin{align}
\ip{\hat\varphi_1^{(0)}\hat\varphi_1^{(1)}} &=-16\alpha\int_0^{1-\sqrt{v_1}}
(s+\frac 32 s^2 +O(s^3)) e^{-8\alpha s} {\rm d}s \nn \\
& +4\int_0^{1-\sqrt{v_1}} (1+2s+3s^2+O(s^3)){e^{-8\alpha s}}{\rm d}s \nn \\
&=-16\alpha\left[\frac 1{(8\alpha)^2}+\frac 3{(8\alpha)^3}
+O(\alpha^{-4})\right] \nn \\
& +4 \left[\frac 1{8\alpha}+\frac 2{(8\alpha)^2}
+ \frac 6{(8\alpha)^3}+O(\alpha^{-4})\right] \nn \\
&=\frac 1{4\alpha} + O(\alpha^{-2}).
\end{align}
Note that the upper bound at $1-\sqrt{v_1}$ has no effect since it
gives a term that decays exponentially with $\alpha$. Thus we find
that the total phase variance is
\begin{equation}
\ip{\phi_{\rm I}^2}=\frac 1{4\alpha}+\frac \tau 2.
\end{equation}
This provides a good verification of the result obtained by the highly
simplified method in \cite{semiclass}.

Note that this result is based on continuing to use the intermediate
phase estimate at the end of the measurement. If we use the phase estimate
$\arg A$ at the end of the measurement we will get a different result, one
that we cannot predict using this approach.

\subsection{Mark II}
If we try to use the same approach for the mark II case it does not seem to be
possible to obtain a consistent result. To illustrate this, we will briefly
outline the derivation.
From \cite{semiclass}, the mark II phase estimate is effectively a time
average of the mark I phase estimates:
\begin{equation}
\phi_{\rm II} \approx \int_0^1 \hat\varphi_t {\rm d}t.
\end{equation}
In order for this to be consistent with the above theory we will
take the average only from time $v_1$, then take the limit of small
$v_1$. In perturbation theory the mark II phase estimate is
\begin{equation}
\phi_{\rm II} = \int_{v_1}^1 [\hat\varphi_t^{(0)}+\alpha\tau\hat
\varphi_t^{(1)}] {\rm d}t.
\end{equation}
We find that the variance is
\begin{equation}
\label{markii}
\ip{\phi_{\rm II}^2}=\int_{v_1}^1 {\rm d}t \int_{v_1}^1 {\rm d}t'
\ip{\hat\varphi_t^{(0)}\hat\varphi_{t'}^{(0)}} + 2\alpha\tau\int_{v_1}^1
{\rm d}t \int_{v_1}^1 {\rm d}t' \ip{\hat\varphi_t^{(0)} \hat\varphi_{t'}^{(1)}}.
\end{equation}

The first term is fairly well behaved, and in the limit of small $v_1$
we get
\begin{equation}
\int_{v_1}^1 {\rm d}t \int_{v_1}^1 {\rm d}t' \ip{\hat\varphi_t^{(0)}
\hat\varphi_{t'}^{(0)}} \approx \frac 1{4\alpha^2}.
\end{equation}
Expanding the second term in equation (\ref{markii}) gives
\begin{align}
\int_{v_1}^1 {\rm d}t \int_{v_1}^1 {\rm d}t'\ip{\hat\varphi_t^{(0)}
\hat\varphi_{t'}^{(1)}} = & \int_{v_1}^1 {\rm d}t \int_{v_1}^1 {\rm d}t'\left[
e^{4\alpha(2\sqrt{v_1}-\sqrt t-\sqrt{t'})}\ip{\hat\varphi_{v_1}^{(0)}
\hat\varphi_{v_1}^{(1)}}\vphantom{\int_{v_1}^{t'}}\right. \nn \\ & \left.
-\int_{v_1}^{t'}\frac {4\alpha}se^{4\alpha(2\sqrt{v_1}-\sqrt t-\sqrt{t'})}
\ip{{\hat\varphi_{v_1}^{(0)2}}} {\rm d}s \right. \nn \\
& \left. -\int_{v_1}^{t'}\frac{4\alpha}s {\rm d}s\int_{v_1}^t\int_{v_1}^s
\frac {e^{4\alpha(\sqrt v-\sqrt t)}}{\sqrt v}{\rm d}W(v)\frac {e^{4\alpha(
\sqrt u-\sqrt{t'})}}{\sqrt u}{\rm d}W(u) \right. && \nn \\ & \left. +\int_{v_1}^
{t}\int_{v_1}^{t'}\frac{e^{4\alpha(\sqrt v-\sqrt t)}}{\sqrt
v}{\rm d}W(v)\frac 2u e^{4\alpha(\sqrt u-\sqrt{t'})}{\rm d}W(u)\right].
\end{align}
Simplifying this, the third and fourth terms cancel, and the first two terms
give
\begin{equation}
\frac {v_1}{4\alpha^2}\ip{\hat\varphi_{v_1}^{(0)}\hat\varphi_{v_1}^{(1)}}
-\frac {\sqrt{v_1}}{2\alpha^2}\ip{{\hat\varphi_{v_1}^{(0)2}}}
+O(\alpha^{-3})
\end{equation}
Unlike the results for the mark I case, all the terms depend on the conditions
at time $v_1$. This means that it is not possible to obtain an unambiguous
result in the limit $v_1\to 0$. Therefore we consider an alternative
approach for estimating the increase in the phase variance due to the time
delay.

\section{Theoretical minimum}
The alternative method of obtaining an estimate of the time delay is to
consider the squeezed state $\ket{\beta,\zeta}$ in the probability
distribution. As was explained above, the excess phase variance due to
the measurement scheme is approximately the phase variance of this
squeezed state.

From \cite{collett}, the phase variance of a squeezed state is given by
\begin{equation}
\ip{\Delta \phi^2}\approx \frac{n_0+1}{4\bar n_{\rm p}^2} + 2 {\rm erfc}
\left( \sqrt{2n_0} \right ),
\end{equation}
where $n_0 = \bar n_{\rm p} e^{2\zeta}$ for real $\zeta$. Here we use the
subscript p to indicate the mean photon number of the squeezed state in
the probability distribution, as opposed to the photon number of the input
state. The average value of $\bar n_{\rm p}$ will be close to the photon number
of the input state.

For states that are significantly less squeezed than optimum, the second
term is negligible and we can omit the term of order $\bar n_{\rm p}^{-2}$. Then
this simplifies to
\begin{equation}
\ip{\Delta \phi^2} \approx \frac{e^{2\zeta}}{4\bar n_{\rm p}}.
\end{equation}
Since $\bar n_{\rm p}$ will be close to the photon number of the input state,
it is reasonable to replace it with $\bar n$.

When we have a delay of $\tau$ in the system, before time $\tau$ we
have no information about the phase of the system to use to adjust the
local oscillator phase. Therefore we must use a heterodyne scheme for this
time period, rapidly varying the local oscillator phase. This means that
$B_{\tau}$ will be equal to zero, and no matter how good the phase estimate
is after time $\tau$, the largest the magnitude of $B_v$ can be made is
$v-\tau$. Then at the end of the measurement, the largest $|B|$ can be is
$1-\tau$, and the largest $|\zeta|$ can be is ${\rm atanh} (1-\tau)$.

The lower limit to the phase variance of $\ket{\beta,\zeta}$ when there is
a time delay of $\tau$ is therefore
\begin{align}
\ip{\Delta \phi^2}_{{\rm min}}& \approx \frac{e^{-2{\rm atanh} (1-\tau)}}
{4\bar n},\\ & \approx \frac{\tau}{8\bar n}.
\end{align}
This is therefore also the lower limit to the introduced phase variance when
there is a time delay of $\tau$. We can expect that the introduced phase
variance will be close to this for states of small intrinsic phase variance,
as there will quickly be very good phase estimates available for the feedback
phase. In addition the time delay must be sufficiently large that the phase
variance given by this expression is significantly above the
introduced phase variance for no time delay.

This result obeys the same scaling law as the result given in
\cite{semiclass}, but it is a factor of four times smaller. Note, however,
that the limit condition for the result given in \cite{semiclass} is that
$\tau \alpha$ is small, whereas the above result should only be accurate
when both $\alpha$ and $\tau$ are reasonably large. The result here also
differs in that it is the limit for the total introduced phase variance,
rather than just the extra phase variance due to the time delay.

\section{Numerical results}
\label{result}
These analytic results were also tested numerically. The numerical
techniques used were similar to those used in reference~\cite{unpub}. Minimum
uncertainty squeezed states were used, with the stochastic differential
equations for the squeezing parameters \cite{rigo}
as given in \cite{unpub}. For all
calculations $2^{20}$ time steps were used, and calculations were performed
with time delays of $2^n$ time steps, where $n$ varies from 0 to 18.

For the first $2^n$ time steps the local oscillator phase was rotated by
$\pi /2$ each step. For the following time steps the data up to the time
step $2^n$ before the current time step was used. For a delay of $2^0=1$
time steps the data from the previous step is used, corresponding to the
technique for no time delay.

Numerical results for four different phase feedback schemes were obtained:

(a) The simplified feedback for mark I and II measurements, where
\begin{equation}
{\rm d} \hat \varphi_v = \frac{I(v) {\rm d}v}{\sqrt v}.
\end{equation}

(b) The unsimplified feedback, where the phase estimate is
\begin{equation}
\hat \varphi(v) = \arg A_v.
\end{equation}

(c) The phase estimate that is intermediate between $\arg A_v$ and the best
phase estimate
\begin{equation}
\hat \varphi(v) = \arg \left( A_v^\epsilon C_v^{1-\epsilon} \right),
\end{equation}

~~~~~where $\epsilon$ is a constant.

(d) The same as in (c), except that the value of $\epsilon$ varies with
time as
\begin{equation}
\epsilon(v) = \frac{v^2-|B_v|^2}{|C_v|}\sqrt{\frac{v}{1-v}}.
\end{equation}

\subsection{Comparison with perturbative theory}
\begin{figure}
\centering
\includegraphics[width=0.6\textwidth]{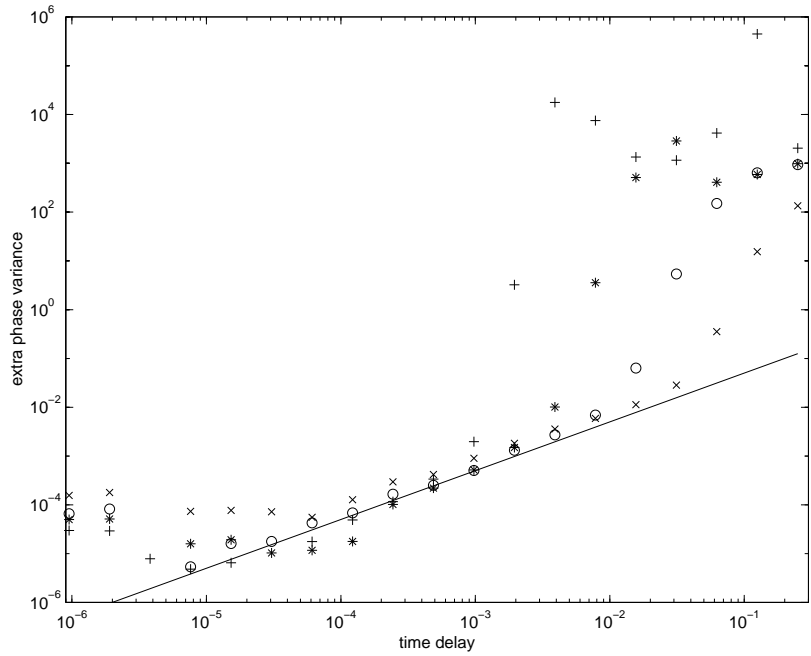}
\caption{The extra phase variance (in the final value of the intermediate
phase estimate for the simplified feedback) due to the time delay plotted
as a function of time delay for four different mean photon numbers. The
data for a mean photon number of 121.590 is shown as crosses, for a
photon number of 1 576.55 as circles, for a photon number of 22 254.8 as
asterisks and for a photon number of 332 067 as plusses. The theoretical
value of ${\tau}/2$ is plotted as the continuous line. }
\label{mark0}
\end{figure}

First we consider the case of simplified feedback, and consider
the variance in the final value of the feedback phase, rather than the phase
of $A$ or $C$. This case was examined in section~\ref{dtontwo}, and the
extra phase variance due to the time delay is $\tau /2$ according to that
analysis. The extra phase variance is plotted for four different mean
photon numbers in figure \ref{mark0}.

In determining the extra phase variance due to the time delay, an estimate
must be made of the phase variance with no time delay. For the results shown
in figure \ref{mark0}, the phase variances are very close for the first six or
so time delays. The estimate of the phase variance with no time delay was
taken to be the minimum of these results.

The theoretical value of $\tau /2$ is also plotted in figure \ref{mark0}.
As can be seen, many of the results are close to the theoretical line for the
intermediate time delays. For small time delays the extra phase variance
due to the time delay is too small a fraction of the total phase variance
for the results to be accurate. The reason why the results deviate from the
theoretical result for large time delays is that this theoretical result
is for the limit of small $\alpha \tau$. Note also that the results for
larger photon numbers deviate from the theoretical result for smaller
$\tau$ than the results for smaller photon numbers. This is also to be
expected from this limit condition.

We also used fitting techniques to determine how closely the numerical
results agree with the theoretical value. A linear fit of the phase
variances against $\tau$ was performed, and data for $\alpha \tau$ above
about $0.3$ was omitted, as this was where the results started to increase
dramatically. The average slope obtained was $0.39\pm 0.06$, in reasonable
agreement with the theoretical value of $0.5$.

\begin{figure}
\centering
\includegraphics[width=0.6\textwidth]{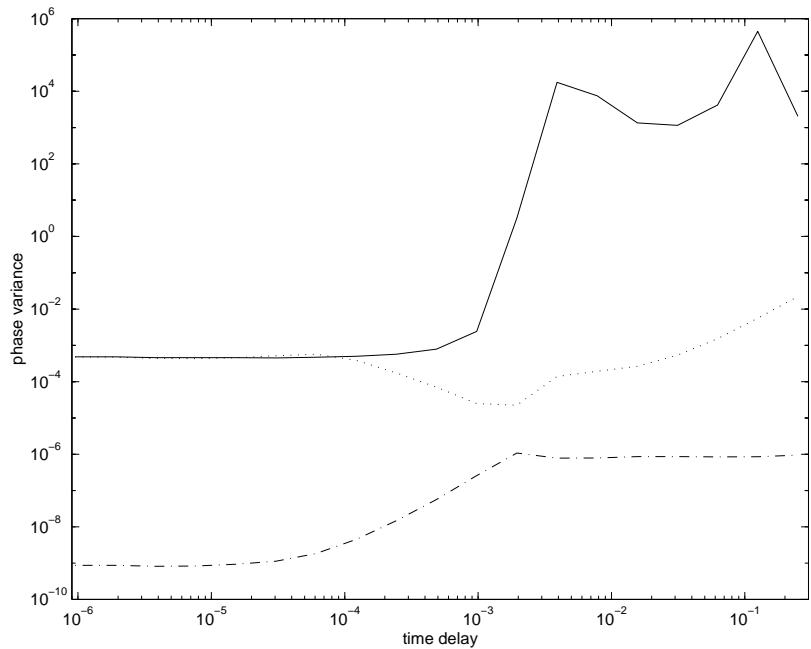}
\caption{The phase variance of three alternative final phase estimates
for simplified feedback with a time delay plotted as a function of time
delay. The results for the final value of the intermediate phase estimate
are plotted as a continuous line, for $\arg A$ as a dotted line, and for
$\arg C$ as a dash-dot line. All results are for a photon number of 332 067.}
\label{all3}
\end{figure}

As was mentioned above, the result for the additional phase variance
due to the time delay is only valid for the variance in the final
value of the phase estimate, which is not the same as $\arg A$ when there
is a time delay. In figure~\ref{all3} we have plotted the variation of the
phase variance with time delay for three alternative final phase estimates,
$\hat \varphi$, $\arg A$ and $\arg C$. This is for a photon number of
approximately 332 000, and is fairly representative of the results for other
photon numbers.

As can be seen, for very small time delays the variances in the
$\hat \varphi$ and $\arg A$ phase estimates are almost identical. As the time
delay is increased, however, the variance of $\hat \varphi$ increases, but
the variance of $\arg A$ {\it decreases}. This is because, as the intermediate
phase estimate gets worse, the value of $|B|$ decreases. This means that
$A$ is closer to $C$, so $\arg A$ is closer to the best phase estimate. Note,
however, that the variance of $\arg A$ rises again, and does not converge to
$\arg C$ for large time delays. This is because $|B|$ does not fall to zero.

\subsection{Comparison with theoretical minimum}
\begin{figure}
\centering
\includegraphics[width=0.6\textwidth]{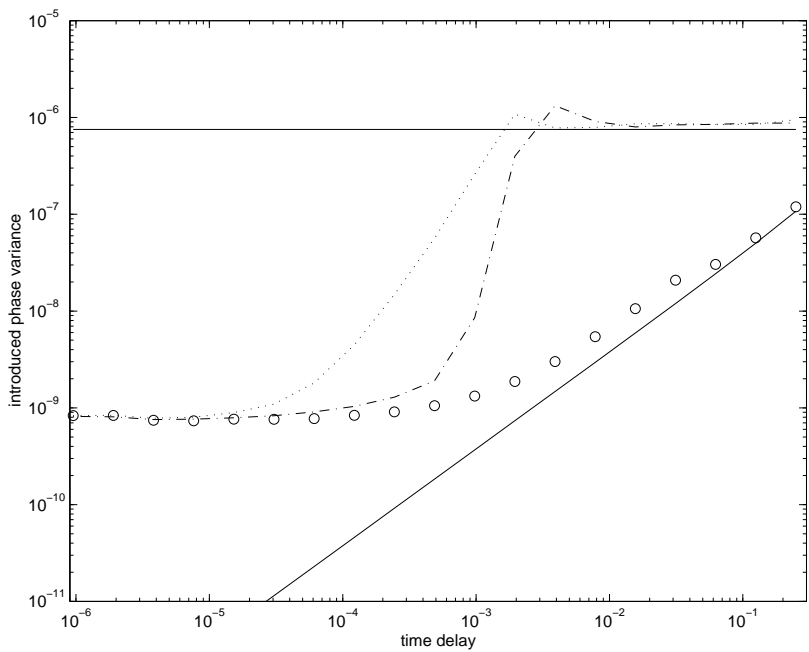}
\caption{The phase variance introduced for three different phase feedback
schemes plotted as a function of time delay. The dotted line is for
simplified feedback, the dash-dot line is for the corrected simplified
feedback, and the circles are for unsimplified $\arg A_v$ feedback. The
best phase estimate $\arg C$ is used in all three cases. The continuous
horizontal line is the phase variance for heterodyne measurements, and
the continuous diagonal line is the theoretical limit. All results are
for a photon number of 332 067.}
\label{limits}
\end{figure}

Now we consider the variance in the phase of $C$. As was explained
above, the theoretical lower limit to the introduced phase variance is
$\tau/(8\bar n)$. We have plotted the introduced variance in the best phase
estimate $\arg C$ and the theoretical limit in figure \ref{limits}. For
additional accuracy we have plotted
\begin{equation}
\frac{e^{-2{\rm atanh}(1-\tau)}}{4\bar n},
\end{equation}
as this will continue to be accurate for time delays that are a large
fraction of 1. This plot is for a photon number of 332 000, and similar
results are obtained for other photon numbers. As can be seen, the phase
variance is well above the theoretical limit. For large time delays the
phase variance approximately converges to the heterodyne phase variance,
also shown in figure \ref{limits}.

The phase variance introduced for mark II measurements with the unsimplified
$\arg A_v$ feedback is also shown in figure~\ref{limits}. The phase variance
introduced for this case increases far more slowly with the time delay,
and for larger time delays it is very close to the theoretical limit. These
results indicate that if there is any significant time delay in the system
the simplified feedback will give a far worse result than using $\arg A_v$.

It is possible to make a correction to the simplified phase feedback scheme
that improves this result somewhat. Many different alternatives were tried,
and the one that gave the best results was 
\begin{equation}
{\rm d} \hat \varphi_v = \frac{I(v) {\rm d}v}{\sqrt {v+\alpha \tau}}.
\end{equation}
This correction is based on the fact that $|A_v|$ is larger than $\sqrt v$
when the phase estimate is worse than $\arg A_v$. (From \cite{semiclass},
the factor of $\sqrt v$ in the simplified feedback comes from a factor of
$|A_v|$.) 

The results for this correction are also shown in figure~\ref{limits}. The
phase variances obtained in this case are significantly below those for
the plain simplified feedback, but are still far above the results for the
unsimplified $\arg A_v$ feedback.

\begin{figure}
\centering
\includegraphics[width=0.6\textwidth]{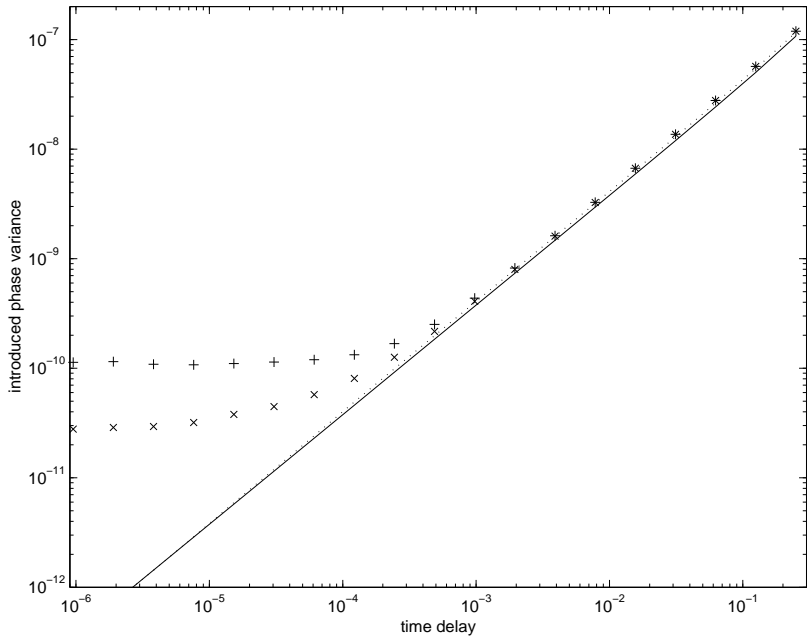}
\caption{The phase variance introduced for better intermediate phase
estimates plotted as a function of time delay. The plusses are for the
constant $\epsilon$ case and the crosses are for the time-dependent
$\epsilon$ case. The theoretical limit estimated using the mean inverse
photon numbers obtained from the time-dependent $\epsilon$ case is plotted
as the dotted line, and the theoretical limit using the input photon
number is shown as the continuous line. All results are for a photon
number of 332 067.}
\label{both50}
\end{figure}

Now we consider the results for better intermediate phase estimates
that are between $\arg A_v$ and $\arg C_v$. The phase variance introduced
for the constant $\epsilon$ case and the theoretical limit are shown in
figure~\ref{both50}. These results are again for a photon number of about
332 000. The results for this case are even closer to the theoretical
limit than those for the mark II case.

The phase variance introduced for the feedback with time dependent
$\epsilon$ is also plotted in figure~\ref{both50}. The results for this case
converge to the theoretical limit at smaller time delays than for the
constant $\epsilon$ case. For the larger time delays the results for
the two cases are about the same, slightly above the theoretical limit.

In both cases we find that for large time delays the phase variance
is still noticeably above the theoretical limit, and that the values
of $B$ obtained are too close to $1-\tau$ to account for this difference.
This difference appears to be due to the approximation that the photon
number of the state $\ket{\beta,\zeta}$ is close to the photon number
of the input state. The average value of this photon number is close to
the photon number of the input state, however each individual value is not
necessarily close to $\bar n$. The expression for the phase variance introduced
depends on the inverse of the photon number, and the average of
an inverse is not necessarily equal to the inverse of an average. The
general expression is
\begin{equation}
\ip{\frac 1n} = \frac 1{\ip n}+\frac{\ip{\Delta n^2}}{\ip n^3}+O(\ip n^{-4}).
\end{equation}
In figure~\ref{both50} we have also plotted the estimated theoretical limit
based on the average of $1/{\bar n_{\rm p}}$ for the data obtained in the time
dependent $\epsilon$ case. Specifically, the expression plotted is
\begin{equation}
\frac 14 \ip{\frac 1{\bar n_{\rm p}}}e^{-2 {\rm atanh}(1-\tau)}.
\end{equation}
As can be seen, the phase variance introduced converges to this far more
closely than to the limit based on the photon number of the input state.

\section{Conclusions}
We have verified that the same result for the increase in phase variance
with time delay for the mark I case with simplified feedback is obtained
by the full perturbation theory calculation as for the highly simplified
calculation in \cite{semiclass}. Our numerical results also verify the
extra phase variance of $\tau/2$ quite accurately.

The result for the increase in the mark I phase variance only holds if
the phase estimate at the end of the measurement is the final value of
the intermediate phase estimate. If the actual value of $\arg A$ is used,
the phase variance decreases for moderate time delays. This is because
the worse intermediate phase estimate reduces the value of $|B|$, making
$\arg A$ closer to the best phase estimate, $\arg C$.

We have also shown that the complete version of the simplified
calculation in \cite{semiclass} to determine the increase in the variance
of the mark II phase estimate does not give convergent results. An
alternative technique shows that the theoretical limit to the introduced
phase variance when there is a time delay of $\tau$ is $\tau/(8\bar n )$.
This is a factor of 4 smaller than the result obtained in
\cite{semiclass}, though it is obtained for different limit conditions.

The introduced phase variance converges to the theoretical limit in the
three different cases with unsimplified feedback considered. (These are
the case with $\arg A_v$ feedback, and the two cases with
$\arg (A_v^\epsilon C_v^{1-\epsilon})$ feedback introduced in
\cite{unpub}.) For the case with simplified feedback, however, the phase
variance is far above the theoretical limit (around 10 times). In this
case, for large time delays the phase variance converges to the heterodyne
phase variance, as the intermediate phase estimate becomes very poor.

It is possible to correct the simplified feedback to reduce the phase variance
greatly, but even this corrected feedback does not give results close to
those for unsimplified $\arg A_v$ feedback. This indicates that if
there is any significant time delay in the system, it is better to use
unsimplified feedback, even though the processing of the data is likely
to introduce a larger time delay. This makes the improved
$\arg (A_v^\epsilon C_v^{1-\epsilon})$ feedback even more attractive, as
one of the main reasons for using the simpler $\arg A_v$ feedback was
that it allows the simplified, analogue feedback circuit to be used.

\end{document}